\documentclass[letterpaper,10pt]{article}

\usepackage[ruled,vlined,linesnumbered]{algorithm2e}
\usepackage{multirow,array,makecell} 
\usepackage{wrapfig,relsize}
\usepackage{authblk}
\usepackage{amsfonts}
\usepackage{fixmath}
\usepackage{setspace}
\usepackage{enumitem}
\usepackage{amsmath}
\usepackage{float}
\usepackage{mathtools,url}
\usepackage{latexsym}
\usepackage{subcaption} 
\usepackage[font={small,sf}]{caption}
\usepackage{tikz}
\usepackage{hyperref}

\makeatletter
\g@addto@macro{\UrlBreaks}{\UrlOrds}
\makeatother

\renewcommand{\footnotesize}{\fontsize{8pt}{10pt}\selectfont}

\usepackage{booktabs}

\newcommand{\OmitText}[1]{ {} }

\usepackage{xspace}

\newcommand{\1}{{\em (i)}}
\newcommand{\2}{{\em (ii)}}
\newcommand{\3}{{\em (iii)}}

\usepackage[all]{hypcap}

\usepackage[capitalise,nameinlink]{cleveref}
\crefname{section}{Sect.}{Sect.}
\Crefname{section}{Section}{Sections}

\usepackage{url}
\makeatletter
\g@addto@macro{\UrlBreaks}{\UrlOrds}
\makeatother
\makeatletter
\def\Url@twoslashes{\mathchar`\/\@ifnextchar/{\kern-.2em}{}}
\g@addto@macro\UrlSpecials{\do\/{\Url@twoslashes}}
\makeatother

\usepackage[framemethod=tikz]{mdframed}
  
\surroundwithmdframed[
  skipabove=0em,
  skipbelow=0ex,
  outerlinewidth=0.4pt,
  innerlinewidth=0.4pt,
  middlelinewidth=1pt,
  middlelinecolor=white,
  bottomline=false,topline=false,rightline=false]{mythm}

\surroundwithmdframed[
  skipabove=1em,
  skipbelow=0ex,
  outerlinewidth=0.4pt,
  innerlinewidth=0.4pt,
  middlelinewidth=1pt,
  middlelinecolor=white,
  bottomline=false,topline=false,rightline=false]{mylem}

\surroundwithmdframed[
  skipabove=0em,
  skipbelow=0ex,
  outerlinewidth=0.4pt,
  innerlinewidth=0.4pt,
  middlelinewidth=1pt,
  middlelinecolor=white,
  bottomline=false,topline=false,rightline=false]{mycor}

\newcounter{myprot}
\newcounter{myalg}
\newcounter{myprop}
\newcounter{myprob}
\newcounter{myexam}
\newcounter{mythm}
\newenvironment{mythm}
{\refstepcounter{mythm}  \noindent \textbf{THEOREM \arabic{mythm}:}\em}
{\vspace{.25em}}

\newcounter{mylem}
\newcounter{mycor}
\newcounter{myobs}
\newcounter{mydef}
\newcounter{myconj}
\newenvironment{myconj}
{\refstepcounter{myconj} \vspace{1em} \noindent \textbf{CONJECTURE \arabic{myconj}:}}
{\vspace{.5em}}

\newenvironment{myproof}
{\noindent \textbf{PROOF:}}
{\vspace{-3ex}\begin{flushright} $\Box$ \end{flushright}\vspace{2ex}}

\begin{document}

\title{\Large Radium: Improving Dynamic PoW Targeting}

\author{George Bissias \\ CICS, UMass Amherst  \\ \texttt{gbiss@cs.umass.edu}}

\maketitle

\begin{abstract}
Most PoW blockchain protocols operate with a simple mechanism whereby a threshold is set for each block and miners generate block hashes until one of those values falls below the threshold. Although largely effective, this mechanism produces blocks at a highly variable rate and also leaves a blockchain susceptible to chain death, i.e. abandonment in the event that the threshold is set too high to attract any miners. A recent innovation called real-time block rate targeting, or RTT, fixes these problems by reducing the target throughout the mining interval. RTT exhibits much less variable block times and even features the ability to fully adjust the target after each block. However, as we show in this paper, RTT also suffers from a critical vulnerability whereby miners deviate form the protocol to increase their profits. We introduce the Radium protocol, which  mitigates this vulnerability in RTT while retaining lower variance block times, responsive target adjustment, and lowering the risk of chain death. We also show that Radium's susceptibility to the doublespend attack and orphaned blocks remains similar to Bitcoin. 
\end{abstract}

\section{Introduction}

To date, the most popular consensus mechanism for public blockchains is proof-of-work (PoW)~\cite{Nakamoto:2009}. Under PoW, a blockchain (or simply \emph{chain}) is secured by compelling participants to provide evidence of wasted computation or \emph{work}. Every unit of work boosts a participant's odds of deciding the content of the next block. If any one individual or group controls the majority of work, then they are capable of deciding the majority of blocks, and it is possible for them to rewrite an arbitrarily long portion of the chain and censor future transactions. Indeed, even if one mining group produces only a significant fraction of the work, then it is still possible for them to rewrite short portions of the blockchain with relatively high probability. This allows for the group to reverse transactions, an activity known as \emph{doublespending}~\cite{Nakamoto:2009,Rosenfeld:2012}. For this reason, blockchain security is intimately tied to the aggregate work required to produce a block. The most popular PoW blockchains have attracted a large number of participants, who collectively expend a great deal of work. Maintaining a consistent level of work is critical both to maintaining  attack protection as well as a stable block time. 

A specific kind of work is required for each blockchain, which we call its \emph{PoW algorithm}. Typically, specialized hardware called an application specific integrated circuit (ASIC) is required for producing work relevant to a given PoW algorithm. As a result, it is common for multiple Blockchains to occupy the same \emph{PoW market} where they compete for security. In order to attract participants to expend work, blockchains offer a subsidy for blocks produced. Recent research (Kwon et al.~\cite{Kwon:2019} and Bissias et al.~\cite{bissias:2019}) has shown that the relative fiat exchange value of these subsidies, across blockchains in the same market, determines the distribution of work. The end result is that participants will frequently shift their work from one chain to another as the value of rewards fluctuate. These fluctuations can be devastating to minority work chains that can experience huge changes in aggregate work in relative terms. In the worst case, a drastic drop in the fiat exchange value for a minority work chain can remove all incentive for miners to allocate it any work. This causes what is called a \emph{chain death spiral}~\cite{phanpp:2017}.  
 
In this paper, we analyze a recently introduced protocol called real-time block rate targeting, or RTT~\cite{Harding:2020}, whose intended purpose is improve responsiveness to changes in hash rate. We show that RTT currently suffers from a vulnerability due to misaligned miner incentives. We then describe a modification to the RTT protocol, which we call  \emph{Radium}, which fixes this problem. Radium retains the benefits of RTT including lower variance block times, a more responsive difficulty adjustment algorithm (DAA), and prevention of chain death. Not previously studied in the RTT paper, we also show that Radium maintains orphan rate and doublespend attack prevention similar to Bitcoin. 

\section{Background and Related Work}
\label{sec:background}

Under PoW, \emph{miners} repeatedly perturb and hash the block header with frequency $h$, which is called the \emph{hash rate}. Miners hope to hash a value that falls below a protocol-defined \emph{target} $G$. When such a value is found, the block is added to the blockchain and the miner receives coins as a reward. Thus coin issuance is tied to block discovery, and so the protocol must adjust $G$ periodically in order to maintain a steady rate of inflation. Closely related to the target and hash rate is \emph{difficulty} $D$, which was shown by Ozisik et al.~\cite{Ozisik:july2017} to be equivalent to the expected number of hashes required to mine a single block whose hash falls below $G$. The authors further showed that $D = S / G$ where $S$ is the size of the hash space. Accordingly, one can equivalently adjust the target by creating an inverse change in difficulty. Indeed, all of the most popular PoW blockchains employ some form of \emph{difficulty adjustment algorithm} or DAA.

\subsection{Difficulty adjustment}

Currently, all DAAs that we are aware of implement a feedback controller, which first forms a statistical estimate of recent hash rates by observing previous block times and then adjusts the difficulty so as to achieve a target block time $\mathcal{T}$. When the difficulty is tuned so that the target block time is achieved, we say that the DAA is \emph{at rest}. For example, every 2016 blocks, Bitcoin (BTC) scales the current difficulty according to 
\begin{equation}
\label{eq:btc_daa}
D' = D \mathcal{T} / \overline{T},
\end{equation}
where $\overline{T}$ is the average actual block time over the previous 2016 blocks. This simple DAA works fairly well for BTC primarily because the blockchain enjoys more than 90\% of the total available SHA256 hash rate. However, for blockchains with a small fraction of the available hash rate, such as Bitcoin Cash (BCH), simple feedback controllers is inadequate~\cite{Stone:2017b}.

\subsection{Conventional PoW mining}
\label{sec:pow_mining}

The distribution of block inter-arrival time under conventional PoW is $\texttt{Expon}(\lambda)$ where $\frac{1}{\lambda} = \mathcal{T}$, the target block time (see Rizun~\cite{Rizun:2016} and Ozisik et al.~\cite{Ozisik:july2017}).
And, as described above, 
\begin{equation}
\label{eq:fundamental}
D = S / G, 
\end{equation}
where $D$ is the expected number of hashes required to mine a block, $G$ is the target, and $S$ is the size of the hash space. 
For fixed hash rate $h$ we have by definition
\begin{equation}
\label{eq:diff_hash}
H = h E[T], 
\end{equation}
where $T$ and $H$ are the actual block time and hashes per block, respectively. In particular, this implies that
\begin{equation}
\label{eq:diff_hash_rest}
D = h \mathcal{T},
\end{equation}
when the DAA is at rest. 

\subsection{RTT mining}
\label{sec:rtt}

Stone~\cite{Stone:2017b} was perhaps the first to suggest the notion of increasing the mining target \emph{during} a single block interval in order to compensate for statistical tail events or a sudden loss of hash rate. Recently, Harding~\cite{Harding:2020} introduced a new PoW consensus mechanism called RTT that leverages this idea. When mining a given block, instead of using a fixed target $G$, RTT varies the target as a function of the time since the last block. This small change is significant because it alters the statistics of the mining process. 

Define the \emph{instantaneous mining rate}, or expected blocks mined per second, for RTT as 
\begin{equation}
\label{eq:lambda_t}
\lambda(t) = a t^{k-1}, 
\end{equation}
with security constant $k$ and tuning constant $a$. Variable $t$ represents the elapsed time since the last block. Let $T$ be a random variable corresponding to the block inter-arrival time. Harding shows that, given instantaneous mining rate $\lambda(t)$, 
\begin{equation}
\label{eq:rtt_dist}
T \sim \texttt{Weibull}(k, a),
\end{equation}
where $T$ has density function
\begin{equation}
\label{eq:weibull_density}
f(t;k, a) = a t^{k-1} e^{-at^k/k},
\end{equation}
distribution function
\begin{equation}
\label{eq:weibull_dist}
F(t;k, a) = 1-e^{-at^k/k},
\end{equation}
and expected value
\begin{equation}
\label{eq:weibull_mean}
E[T] = \left(\frac{k}{a}\right)^{1/k} \Gamma \left( 1 + \frac{1}{k} \right).
\end{equation}
From Equation~\ref{eq:weibull_mean}, it is evident that, when targeting a block in expected time $\mathcal{T}$, the constant $a$ should be defined as
\begin{equation}
\label{eq:a}
a = k \left[\frac{\Gamma \left(1+ \frac{1}{k}\right)}{\mathcal{T}} \right]^k
\end{equation}

RTT is designed to maintain compatibility with Bitcoin, which requires RTT to maintain conventional mining targets on the blockchain: $G_i$ for each block $i$. This has the primary benefit of maintaining blockchain continuity before and after upgrade and the ancillary benefit of allowing for conventional difficulty adjustment. From conventional target $G_i$ and instantaneous mining rate $\lambda_i(t)$, RTT requires a \emph{subtarget} $g_n(t)$ such that the expected mining time for RTT under $g_n(t)$ is equal to target mining time $\mathcal{T}$. To mine block $i$,  miners must find a block at some elapsed time $t$ whose hash falls below $g_i(t)$.

Under conventional PoW, $G_i$ implies an instantaneous block mining rate of 
\begin{equation}
\label{eq:conv_mining_rate}
\lambda = 1/\mathcal{T}
\end{equation} 
blocks per second. Accordingly, the sub-target $g_i(t)$ is defined as
\begin{equation}
\label{eq:subtarget}
g_i(t) = G_i \frac{\lambda_i(t)}{\lambda},
\end{equation}
which can be interpreted as normalizing $G_n$ according to the variable block production rate of RTT.

\section{Future Mining Attack on RTT}
\label{sec:future_mining_attack}

In this section we describe a \emph{future mining} attack on RTT. Miner $A$, having fraction $q$ of the total hash rate, chooses a future time $t^*$ and allocates all of his hash rate to finding a block that meets sub-target $g(t^*)$ until $t^*$ has expired. There are three mutually exclusive outcomes: \1 $A$ mines a block prior to $t^*$; \2 the remaining miners $M$, having fraction $1-q$ of the hash rate, mine a block prior $t^*$; or \3 a block is first mined after $t^*$.
For outcome \3, we assume that $A$ will revert to protocol-compliant mining, i.e. $A$ will mine to actual time $t$ provided that $t \geq t^*$.
In this section we identify a value of $t^*$ such that the probability of $A$ mining a block before $M$ exceeds $q$, which is his fair share. 

\subsection{Attacker expected block time}
\label{sec:attack_expon}

Let $T_A$ and $T_M$ denote statistics corresponding to the time required for $A$ and $M$, respectively, to mine their next blocks. And let $p(t^*)$ denote the probability that $A$ mines a block before $M$ when his future mining time is $t^*$. Note that
\begin{equation}
\label{eq:future_mine_prob}
p(t^*) \geq P(T_A < t^*, T_M > t^*) = P(T_A < t^*) P(T_M > t^*).
\end{equation}
Because $A$ mines with a fixed target $g(t^*)$ for each block, $T_A$ is exponentially distributed (as described in Section~\ref{sec:background}). 

It is apparent from Equations~\ref{eq:fundamental}--\ref{eq:diff_hash_rest} that 
if $D$ is initially tuned for hash rate $h$ and target $G$, but all miners instead mine according to the sub-target at time $t^*$, then  they would expect a block to arrive in time 
\begin{equation}
E[T] = \frac{S}{h g(t^*)} = \frac{G}{g(t^*)} \frac{S}{h G} = \frac{G}{g(t^*)} \mathcal{T}.
\end{equation}

For miner $A$, the expected block time is scaled by his fraction of the total hash rate $q$. Therefore, the expected block time for $A$ is 
given by
\begin{equation}
\label{eq:future_mine_exp_time}
E[T_A] = \frac{S}{q h g(t^*)} =  \frac{G}{q g(t^*)} \frac{S}{h G} =  \frac{G}{q g(t^*)}  \mathcal{T}.
\end{equation}

\subsection{Compliant expected block time}

\begin{figure}
  \centering
  \includegraphics[width=0.8\linewidth]{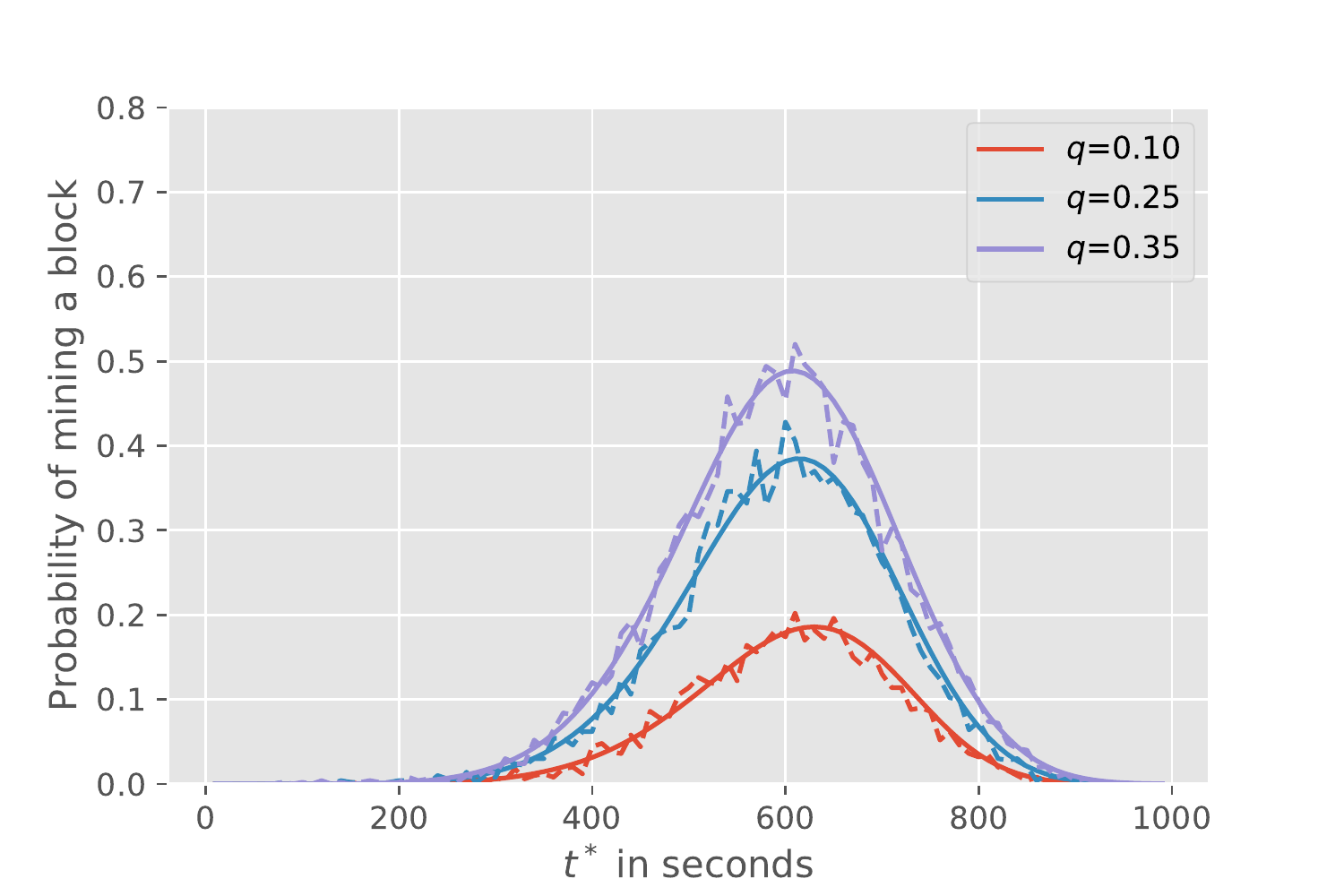}
  \caption{Probability of successful future mining attack for various attacker shares of total hash rate (each curve) and future mining times (independent axis). Solid curves indicate theoretical probabilities and dashed curves indicate the results of a mining simulation with 500 trials per point, per curve.}
  \label{fig:future_mine_attack}
\end{figure}

From the discussion in Section~\ref{sec:rtt}, it is clear that $T_M$ has Weibull distribution since miners $M$ are assumed to be compliant. But they are missing fraction $q$ of the hash rate, which we must account for in determining their expected mining time. It is difficult to reason directly about how Weibull mining time changes with a loss of hash rate, but straightforward to reason about the effect of such a change under conventional PoW. 

Let $T_M'$ be a random variable representing the mining time for miners $M$ having fraction $p = 1-q$ of the total hash rate under the assumption that they use conventional PoW, i.e. mining to fixed target $G$ for each block. Reasoning similarly to Equation~\ref{eq:future_mine_exp_time}, we have 
\begin{equation}
E[T_M'] = \frac{S}{phG} = \frac{1}{p} \frac{S}{hG} = \frac{\mathcal{T}}{p}.
\end{equation}
Target $G$ is a conventional PoW target, so we reason that mining in RTT with fraction $p$ of the total hash rate is equivalent to \emph{all miners} mining against initial target $p G$. The sub-target is related to $G$ by $g(t) = G \lambda(t) / \lambda$. This implies that a sub-target adjusted for hash rate $p$ would be 
\begin{equation}
g_M(t) = p G \frac{\lambda(t)}{\lambda} = G \frac{p \lambda(t)}{\lambda} = G \frac{p a t^{k-1}}{\lambda} = G \frac{\lambda_M(t)}{\lambda},
\end{equation}
where $\lambda_M(t) = p a t^{k-1}$. Thus, according to Equations~\ref{eq:lambda_t} and~\ref{eq:rtt_dist},
\begin{equation}
\label{eq:scaled_weibull}
T_M \sim \texttt{Weibull}(k, pa). 
\end{equation}
Finally, we can produce the bound for $p(t^*)$ in Equation~\ref{eq:future_mine_prob} by multiplying the CDF for $T_A$, evaluated at $t^*$ by the inverse-CDF for $T_M$, evaluated at $t^*$.

Figure~\ref{fig:future_mine_attack} shows the associated probability of mining a block when future mining for many possible future times $t^*$. Each curve corresponds to a different fraction of the total hash rate for $A$. Solid lines are those predicted by the bound in Equation~\ref{eq:future_mine_prob} and dashed lines are the results of a mining simulation. The plot shows that for each hash rate, there exists a regime of values for $t^*$ where the probability of mining a block is greater than the \emph{fair} probability (equivalent to $A$'s share of the hash rate). 

\section{Defacto Future Mining in RTT}

In Section~\ref{sec:future_mining_attack} we showed that the RTT protocol is vulnerable to future mining. This attack arrises because miners are incentivized to mine to a fixed target in the future rather than adhere to the dynamic target established by the protocol. Therefore it seems that future mining is inevitable in RTT. Yet it seems possible that RTT could still be fair for all miners if they all future mine so as to maximize their individual profit. In this section, we show that there is a unique Nash equilibrium future mining time $\tau$, depending on the chain's profitability relative to other chains, to which all miners will mine. Once this time expires, the Nash equilibrium behavior is to mine according to the compliant RTT sub-target.

\subsection{Block preemption}

Future mining involves mining to a target $g(t^*)$ corresponding to a future time $t^*$. Because it is a probabilistic process, the miner will fail to mine a block in time $t^*$ with some frequency. At this point, he must choose a new time $t^*|t$, given that $t$ seconds have passed. This process continues until a block is mined. 

There exists a risk / reward tradeoff in future mining related to the fact that a miner cannot release a block mined at a future time until that time arrives. Thus, if miner $M$ mines to future time $t^*$, then the remaining miners $M_{-}$ can mine to time $t^* - \epsilon$, $\epsilon > 0$, and all blocks mined by $M$ can be preempted by any block mined by $M_{-}$ prior to $t^*$. We call this process \emph{block preemption}. 

\subsection{Game theoretical results}

\subsubsection{Assumptions}

In our game theoretical model, we assume all miners follow a strategy where they will future mine whenever it is possible to do so without being preempted. 

Miners typically have a choice where they direct their hash rate. Let $r$ denote the (fixed) prevailing \emph{reward rate}, which is the amount of fiat that can be gained per hash when miners mine on a competing blockchain. Furthermore, let $R(t^*)$ denote the reward rate when future mining to time $t^*$ on the RTT chain. We assume that a rational miner will choose to direct all of his hash rate to a competing chain whenever $R(t^*) < r$. Thus, we imagine that miners choose $t^*$ so as to maximize $R(t^*)$. We begin with the following result showing that the Nash equilibrium for $t^*$ as a function of $r$.

\bigskip

\begin{mythm}
\label{thm:defacto_nash_1}
There exists a unique Nash equilibrium for initial future mining time $t^* = \tau$, where $R(\tau) = r$.
\end{mythm}

\begin{myproof}
Consider the best response $t^*(M)$ for miner $M$ given a known choice for $t^*(M_{-}) > \tau$ for the remaining miners $M_{-}$. By choosing $t^*(M) = t(M_{-})^* -\epsilon$, for an infinitesimally small $\epsilon > 0$, $M$ can be certain that his blocks will preempt those of $M_{-}$ (ignoring block propagation delay). Moreover, $M$ mines at effectively the same difficulty as $M_{-}$ for each $t(M)^*$. Therefore, the expected profit per hash for $M$ is strictly superior to that of $M_{-}$. Now consider the best response for $M$ when $M_{-}$ chooses $\textbf{t}(M_{-})^* = \tau$. If $M$ chooses to mine to time $\tau - \epsilon$, then he will be mining at a loss relative to  prevailing reward rate $r$. On the other hand, if $M$ mines to future time $\tau + \epsilon$, then his blocks will be preempted by any blocks mined by $M_{-}$. Therefore, the best response for $M$ is also to future mine to time $\tau$. It follows that $\tau$ is a Nash equilibrium for $t^*$. 
\end{myproof}

Having established an equilibrium for the first future mining time, we turn now to subsequent times, which will be targeted in the event that no block is found by time $t^* = \tau$. Somewhat surprisingly, the Nash equilibrium for $t^*|t$ turns out to be equal to the current time $t$ itself.

\bigskip

\begin{mythm}
\label{thm:defacto_nash_2}
For any $t > \tau$, $t^*|t = t$ is a unique Nash equilibrium.
\end{mythm}

\begin{myproof}
Consider the best response for miner $M$ given that $M_{-}$ is mining to time $t^*(M_{-})|t = t$, when $t > \tau$.
$M$ certainly wishes to mine on the RTT chain since $R(t) > r$ for $t > \tau$. But because $t$ is not in the future, it is not possible for $M$ to mine to a slightly earlier time. And if $M$ was to mine to a future time $t + \epsilon$, for $\epsilon > 0$, then it would be possible for his blocks to be preempted. Therefore, the best response for $M$ is to also mine to $t^*(M)|t = t$. It follows that $t^*|t = t$ is a Nash equilibrium.
\end{myproof}

Together, Theorems~\ref{thm:defacto_nash_1} and~\ref{thm:defacto_nash_2} show that all RTT miners will future mine to time $\tau$ until the actual time $t$ exceeds $\tau$ at which point they will revert to (compliant) mining against target $g(t)$. There are three major issues with this behavior. First, at time $\tau$, there is a significant chance that multiple miners will have already future mined a block, creating a block race and, inevitably, a higher block orphan rate. Second, suppose that on average fraction $\alpha$ of the total hash rate is devoted by all miners to future mining to $\tau$, with the remaining $1-\alpha$ being devoted to mining via dynamic target after $\tau$. By future mining to $\tau - \epsilon$, attacker $A$ can preempt any block future mined by other miners. So at arbitrarily small cost, $A$ eliminates orphan risk for fraction $\alpha$ of his blocks. This makes both censoring and doublespending transactions easier for the attacker. A third drawback to defacto future mining is that target $G$ no longer quantifies the actual security applied to the chain. Under conventional PoW, Equation~\ref{eq:fundamental} can be used to determine $D$, which is equivalent to the expected number of hashes performed per block, a proxy for blockchain security. However, given defacto future mining on RTT, the target overshoots the actual hashes per block because miners never mine to a target greater than $g(\tau) < G$.

\section{Radium Protocol}

In this section we present the Radium protocol, which is an extension of RTT. The primary difference is that, in Radium, block reward is also scaled with inter-block time. This causes the reward per hash to remain uniform for a given block (much like conventional PoW), eliminating the profitability of future mining.

\subsection{Mining} 

Radium targets a 600 second average block time like bitcoin, i.e. $\mathcal{T} = 600$. Like RTT, the mining target at each second $t$ after the last block is given by sub-target $g_k(t)$ where $k=2$. For a given target $G$, and combining Equations~\ref{eq:a}--\ref{eq:subtarget}, we have
\begin{equation}
g_k(t) = G \frac{\lambda(t)}{\lambda} = G \frac{at^{k-1}}{1/\mathcal{T}} = G \mathcal{T} t^{k-1} k \left[ \frac{\Gamma(1 + \frac{1}{k})}{\mathcal{T}} \right]^k = k G \Gamma\left(1 + \frac{1}{k}\right)^k \frac{t^{k-1}}{\mathcal{T}^{k-1}}.
\end{equation}
Radium can use any PoW algorithm, for example SHA256. Mined blocks are rapidly propagated header-first to all other miners. If the timestamp of the block is drastically different than the time on the recipient's machine, then it is discarded. In practice the time difference can be as little as the maximum expected header propagation delay if miners use NTP to coordinate clocks. The use of NTP in the Bitcoin network has been discussed as a possibility in the past~\cite{BIP:2014}. Note that NTP synchronized clocks are to aid compliant miners. \textbf{Radium does not rely on dishonest miners reporting accurate time.}

\subsection{Rewards} 
\label{sec:radium_rewards}

Let $d(t) = S / g_k(t)$ be the \emph{sub-difficulty}, where $S$ is the size of the hash space.
Define a new reward function $r(t) = C \frac{d(t)}{d(\mathcal{T})}$ for a block mined at elapsed time $t$. By construction, $r(t)$ will pay out exactly $C$ coins when a block is mined in target time, $\mathcal{T}$ seconds. And it will pay out more or less than that if the block is mined, respectively, sooner or later. The appeal of using $r(t)$ is that the expected reward-per-hash for mining at any given sub-target $g_k(t^*)$ is constant: $\frac{C}{d(T)}$. Thus, this new reward function will serve to disincentivize future mining.

One of the features of the RTT protocol is that the risk of entering a chain death spiral is eliminated because the difficulty will eventually approach zero as time progresses. The same is true for Radium, except that the reward payout also approaches zero. Thus, in relative terms, the reward per hash never increases in Radium. However, chain death remains highly unlikely because the difficulty will eventually become so low that a block can be easily mined on a single CPU with minimal effort. 

\subsection{Difficulty adjustment} 

\begin{figure}
  \centering
  \includegraphics[width=0.7\linewidth]{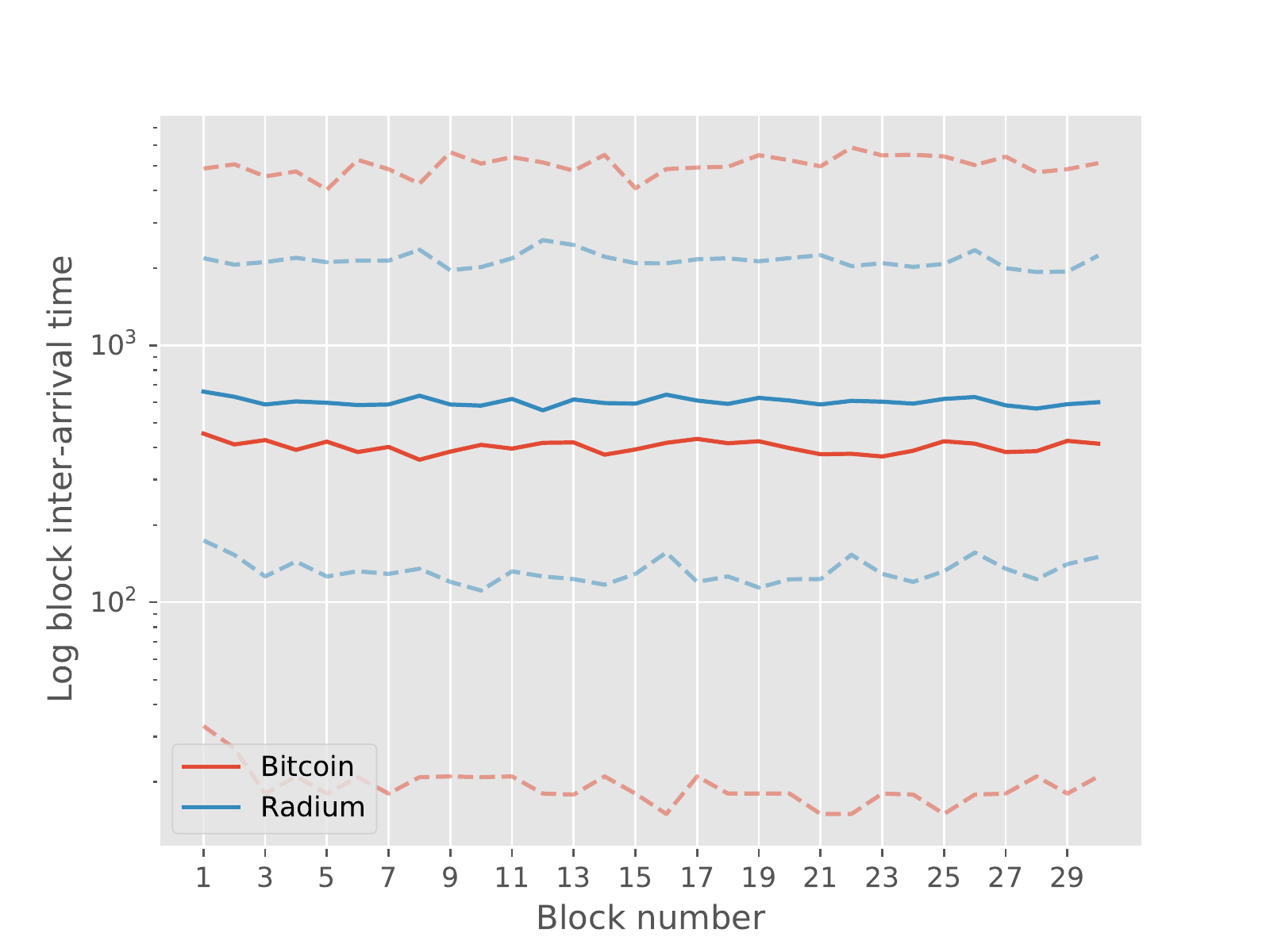}
  \caption{Median block times for Bitcoin (red) and Radium (blue) across 1000 trials of a 30-block simulation (5th and 95th percentiles shown as dashed lines). After each block in the simulation, the difficulty is adjusted according to Equation~\ref{eq:btc_daa} for Bitcoin and Equation~\ref{eq:radium_daa} for Radium, with $\overline{T}$ being the mean of the previous two block times.}
  \label{fig:daa}
\end{figure}

Radium uses feedback control to adjust its difficulty in much the same way as conventional PoW protocols. In particular, it adopts the same mechanism used by RTT, which we describe and refine presently. 

Mining amounts to successive draws of random variable $T$ (representing block time) from a given distribution, while difficulty adjustment involves estimating the current scale of the block time distribution from $T$ and moving that scale closer to the ideal. Therefore, difficulty adjustment is essentially parameter estimation from sample $T$. Rather than directly estimating the scale of Weibull distributed $T$, Harding~\cite{Harding:2020} opts to transform $T$ to an exponentially distributed random variable $T'$ and estimate its scale instead. Because this transformation amplifies distortions due to hash rate fluctuations, he finds that a single sample is often sufficient to accurately update the difficulty. 

\bigskip

\begin{mythm}
A block with inter-arrival time $T$ mined under the Radium protocol with target $G$ would have inter-arrival time $T' = \frac{a T^k \mathcal{T}}{k}$ if mined under conventional PoW with target $G$.
\end{mythm}

\begin{myproof}
Let $T$ be a random variable drawn from $\texttt{Weibull}(k, a)$ and having CDF $F(t; k, a)$ as defined in Equation~\ref{eq:weibull_dist}. The probability integral transform (PIT) dictates that random variable $U = F(T; k,a)$ has distribution $\texttt{Uniform}(0,1)$. Now define $H(x; \lambda) = 1 - e^{-\lambda x}$, the CDF of the distribution $\texttt{Expon}(\lambda)$, where $\lambda$ is defined in Equation~\ref{eq:conv_mining_rate}. We recover $T' \sim \texttt{Expon}(\lambda)$ by applying the PIT in reverse using $H^{-1}$: 
\begin{equation}
\begin{array}{rcl}
T' &=& H^{-1}(U; \lambda) \\
&=& - \frac{\ln(1-U)}{\lambda} \\
&=& - \frac{\ln(1-F(T; k,a))}{\lambda} \\
&=& - \frac{\ln(1- 1 - e^{a T^k / k})}{\lambda} \\
&=& \frac{a T^k \mathcal{T}}{k}.
\end{array}
\end{equation}
\end{myproof}

Suppose that miners currently operate with hash rate $h$ and that for block $i$, it happens that $T_i' \neq \mathcal{T}$. Being exponential, the actual inter-arrival time for block $i$, $T_i'$, is an unbiased estimator of its expected value, i.e. $T_i' \approx E[T_i'] = \frac{1}{\lambda_i}$. Now suppose that $T_i' \neq \mathcal{T}$. We seek to adjust $G_{i+1}$ so that $T_{i+1}' = \mathcal{T}$. Note that, because $D_i$ represents the expected number of hashes required to mine a block, $T_i' \approx 1 / \lambda_i = D_i / h$. And, according to Equations~\ref{eq:conv_mining_rate} and~\ref{eq:diff_hash_rest}, 
\begin{equation}
\frac{1}{\lambda_i} = \mathcal{T} = \frac{D_i}{h} = \frac{S}{G_i h} 
\end{equation}
when the DAA is at rest, which implies that it is necessary to revise $G_{i+1}$ so that $\mathcal{T} = S / (G_{i+1} h)$. We have,
\begin{equation}
\begin{array}{rccl}
& T_i' &\approx& \frac{D_i}{h} \\ 
\Rightarrow & \frac{a t^k}{k} \mathcal{T} &\approx& \frac{D_i}{ h}. \\
\Rightarrow & \mathcal{T} &\approx& \frac{k}{a t^k} \frac{D_i}{h} \\
& &=& \frac{D_i}{(a t^k/k) h}.
\end{array}
\end{equation}
It follows that an update for the difficulty, based on the mean of the previous $n$ block inter-arrival times $\overline{T}$, is given by
\begin{equation}
\label{eq:radium_daa}
D_{i+1} = \frac{k}{a \overline{T}^k} D_i.
\end{equation}

\begin{table}[h!]
\centering
\begin{tabular}{l c c c}
\toprule
{\bf Statistic } ~~&~~ {\bf 5th percentile} ~~&~~ {\bf median} ~~&~~ {\bf 95th percentile} \\
\cmidrule(lr){1-4} \cmidrule(lr){1-4}
Two-sample Bitcoin ~~&~~ 18s ~~&~~ 410s ~~&~~ 4994s \\
Bitcoin Ideal ~~&~~ 31s ~~&~~ 416s ~~&~~ 1797s   \\
Two-sample Radium ~~&~~ 129s ~~&~~ 599s ~~&~~ 2114s \\
\bottomrule
\end{tabular}
\medskip
\caption{Block time statistics for simulations of both Bitcoin and Radium when the difficulty is adjusted every block based on the previous two block times. Statistics of the distribution $\texttt{Expon}(\mathcal{T})$, representing the best possible variability for Bitcoin, are provided for comparison. Each statistic reported is itself the result of the median over the 30-block simulation, with 1000 trials performed per block.}
\label{table:daa}
\end{table}

\subsection{Block time simulation}

We ran a mining simulation in both Bitcoin and Radium that updated the difficulty in each according to Equation~\ref{eq:btc_daa} and Equation~\ref{eq:radium_daa}, respectively, where $\overline{T}$ was the mean of the last two block times. Each trial of the simulation ran for 30 consecutive blocks for 1000 trials total. On a log scale, Figure~\ref{fig:daa} shows the median and 5th and 95th percentiles for Bitcoin and Radium. Not surprisingly, Bitcoin block times (red) show large variability. However, Radium (blue) shows much better concentration of block times around the target of 600s. 

Table~\ref{table:daa} quantifies the median (across the 30 blocks) of medians and 5th and 95th percentiles (each across the 1000 trials). It includes the same statistics for distribution $\texttt{Expon}(\mathcal{T})$, which corresponds to the best possible variability for Bitcoin, when the ideal difficulty is known. The table shows that the median block time for Radium is much closer to the target time of 600s than either two-sample Bitcoin or the ideal. Also, the 5th percentile for two-sample Radium avoids producing extremely early blocks. Finally, the 95th percentile of block times for Radium stays within 18\% of the 95th percentile of Bitcoin Ideal. In contrast, two-sample Bitcoin adjustment is almost 3 times the ideal. \emph{Overall, we find that two-sample Radium difficulty adjustment performs nearly as well as the best possible Bitcoin difficulty adjustment algorithm.}

\subsection{Reduction in block time variance}
\label{sec:variance}

A major feature of RTT is that its Weibull distributed block times have lower variance than exponentially distributed block times under conventional PoW. This affords RTT, and Radium by proxy, with more reliable block inter-arrival times. In this section, we calculate the variances of the Radium block time and compare it to that of the Bitcoin protocol.

We compare the variance in block time for Radium relative to the Bitcoin protocol when the expected block times are both equal to $\mathcal{T}$. To that end, let $X$ and $Y_k$ be random variables representing the block times for Bitcoin and Radium (for given $k$), respectively. Section~\ref{sec:pow_mining} explains that $X$ is exponentially distributed with mean $\mathcal{T} = 1 / \lambda$. It is well known that the variance of the exponential distribution is equal to the square of its mean, i.e. $Var_{\mathcal{T}}[X] = \mathcal{T}^2$. On the other hand, Section~\ref{sec:rtt} explained that block times have distribution $\texttt{Weibull}(k, a)$, with $a$ is defined in Equation~\ref{eq:a}. For the mean of the Weibull distribution we have $E[Y_k] = \gamma \Gamma(1 + 1/k)$, where 
\begin{equation}
\gamma = \left( \frac{k}{a} \right)^{1/k} = \frac{\mathcal{T}}{\Gamma(1 + 1/k)}.
\end{equation}
Note that, by construction, $E[Y_k] = \mathcal{T}$. Next, the variance of $Y$ is given by 
\begin{equation}
Var[Y_k] =\gamma^2 \left[\Gamma(1 + 2/k) - \Gamma(1 + 1/k)^2 \right].
\end{equation}
Thus, when blocks are expected every $\mathcal{T}$ seconds, the variance is 
\begin{equation}
Var[Y_k] = \left(\frac{\mathcal{T}}{\Gamma(1 + 1/k)} \right)^2 \left[\Gamma(1 + 2/k) - \Gamma(1 + 1/k)^2 \right].
\end{equation}
Finally, the \emph{improvement} in variance when adopting Radium over Bitcoin is equal to
\begin{equation}
\frac{Var[Y_k]}{Var[X]} = \left(\frac{1}{\Gamma(1 + 1/k)} \right)^2 \left[\Gamma(1 + 2/k) - \Gamma(1 + 1/k)^2 \right] = \frac{\Gamma(1 + 2/k)}{\Gamma(1 + 1/k)^2} - 1.
\end{equation}
In particular, the improvement for $k=2$ becomes
\begin{equation}
\frac{Var[Y_2]}{Var[X]} = \frac{\Gamma(2)}{\Gamma(3/2)^2} - 1 = \frac{4}{\pi} - 1 \approx 0.27.
\end{equation}

\subsection{Orphan rate}

Perhaps the only drawback of more reliable block times is an increase in the rate that \emph{orphan} blocks are produced. An orphan occurs when two viable blocks are produced at roughly the same time. Miners will ultimately settle on one to form the tip of the blockchain, while the other will be discarded. 

Fortunately, the orphan rate observed under the Radium protocol is not expected to be much worse than the orphan rate observed for Bitcoin. To demonstrate this, we ran a mining simulation of both the Bitcoin and Radium protocols for more than 850,000 blocks each. Any time two blocks were generated within the same three second time period, we incremented the orphan counter. A three second interval was chosen because it represents a realistic delay in today's Bitcoin network~\cite{Nagayama:2019}. Our simulation showed that Bitcoin is expected to experience orphans approximately 0.22\% of the time while Radium is expected to experience orphans about 0.36\% of the time. Thus, Radium's orphan rate is approximately 63\% greater than that of Bitcoin; yet the absolute rate remains low.

\section{Radium Security Analysis}

In this section, we analyze various aspects of Radium protocol security. Our primary focus is on incentivizing protocol compliance as well as mitigating the effects of common attacks on PoW blockchains.

\subsection{Reward function exploitation}

A major concern with using dynamic reward function $r(t)$ (defined in Section~\ref{sec:radium_rewards}) over a constant reward function is that a miner with a large amount of hash rate might suddenly switch from one blockchain (say BTC) over to the Radium chain and mine a block at a very high difficulty so as to gain excessive reward. We call this behavior \emph{switch-mining}. The following argument attempts to show the conditions under which miners can and cannot profit in this fashion. We find that when $k \leq 2$, there exists no advantage to switch-mining between Radium and another chain. 

Suppose that for block 1, a miner from chain $X$ suddenly increases the hash rate on the Radium chain by multiple $x > 1$. Equation~\ref{eq:scaled_weibull} shows that the resulting block time distribution is $\texttt{Weibull}\left(k, x a \right)$. Combining Equations~\ref{eq:a} and~\ref{eq:weibull_mean} it follows that the expected block time $E[T_1]$ will be 
\begin{equation}
E[T_1] = \Gamma(1 + 1/k) \left( \frac{k}{x a} \right)^{1/k} = \mathcal{T} x^{-1/k}.
\end{equation}
Thus, the expected reward $E[R_1]$ amounts to 
\begin{equation}
E[R_1] = C \frac{d(\mathcal{T} x^{-1/k})}{d(\mathcal{T})} = C \frac{g(\mathcal{T})}{g(\mathcal{T} x^{-1/k})} = C \frac{\mathcal{T}^{k-1}}{(\mathcal{T} x^{-1/k})^{k-1}} = C x^{(k-1)/k}.
\end{equation}
Of course, the DAA will respond by adjusting the target so that the increased hash rate yields a block in time $\mathcal{T}$. Next, suppose that, for block 2, the miners withdraw their hash rate. The affect of this withdrawal is inverse-symmetric to the affect of the increase; it follows by substituting y = 1/x in the equations above. We have $E[T_2] = \mathcal{T} x^{1/k}$
and $E[R_2] = C x^{-(k-1)/k}$.

\begin{figure}
  \centering
  \includegraphics[width=0.7\linewidth]{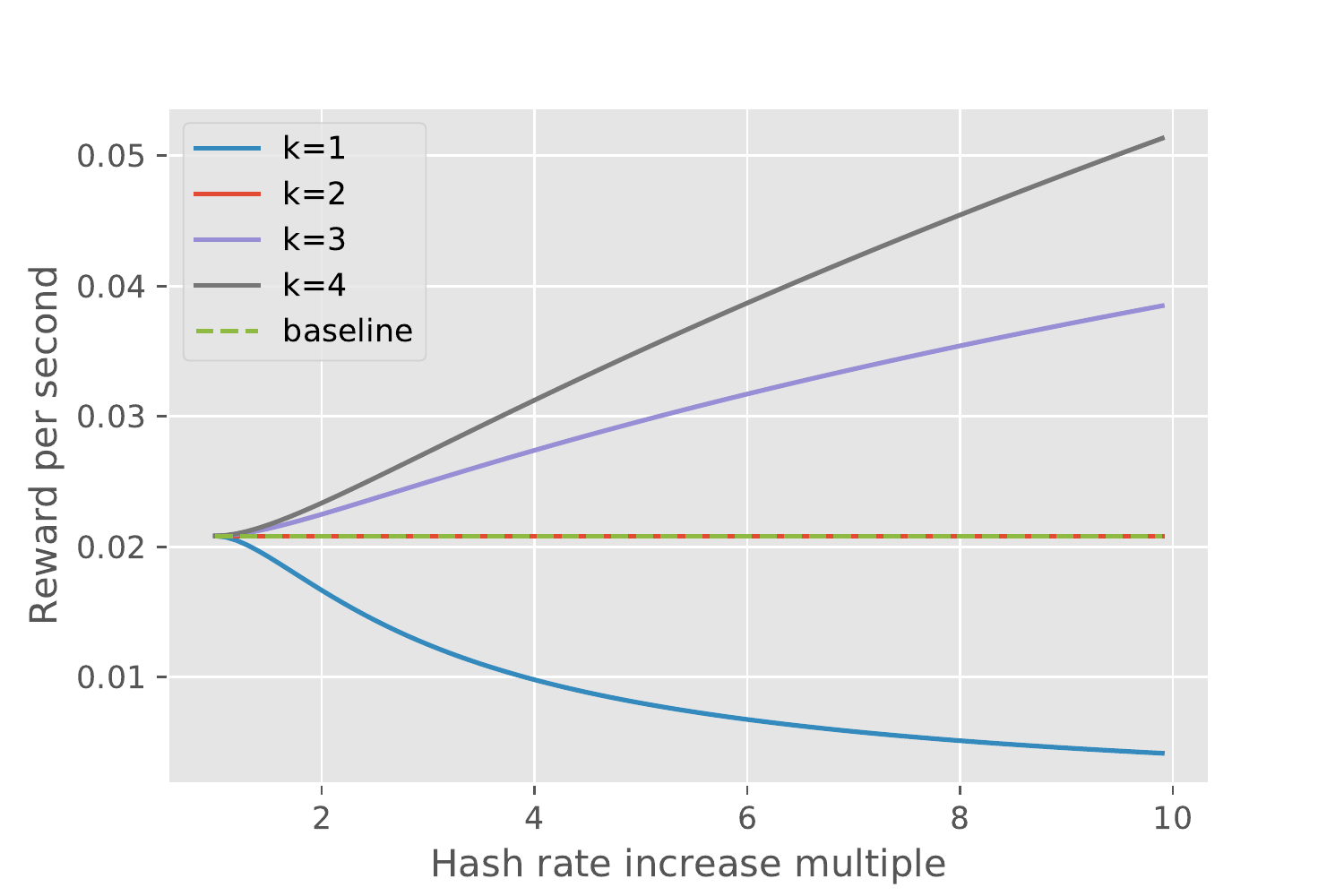}
  \caption{Expected reward per second for a network of miners who switch-mine between Radium and another coin such as Bitcoin. Each curve corresponds to a different value of $k$ from Equation~\ref{eq:a}. The dashed line indicates reward per second if miners do not switch.}
  \label{fig:reward_per_second}
\end{figure}

Figure~\ref{fig:reward_per_second} shows the reward-per-second as a function of hash rate increase multiples $x$ for various values of $k$ where $C = 12.5$. Because the attack takes two blocks to carry out, we measure aggregate reward over both blocks. The results are compared to \emph{baseline}, where hash rate does not fluctuate. We can see from the figure that it is indeed possible for miners to profit, per-unit-hash, for values of $k$ exceeding 2. However, when $k=1$, miners actually lose profit, and for $k=2$, there is no change in profitability.

\subsection{Doublespend attack susceptibility}

Bissias and Levine~\cite{Bissias:2020} argue that high variance is at the core of two of the most fundamental attacks on PoW blockchains: the doublespend and selfish mining. Their Bobtail protocol demonstrates that a lower variance block time can substantially mitigate both attacks. We can compare Radium's improvement in variance over Bitcoin (see Section~\ref{sec:variance}) to that of Bobtail over Bitcoin.

\begin{figure}
  \centering
  \includegraphics[width=0.7\linewidth]{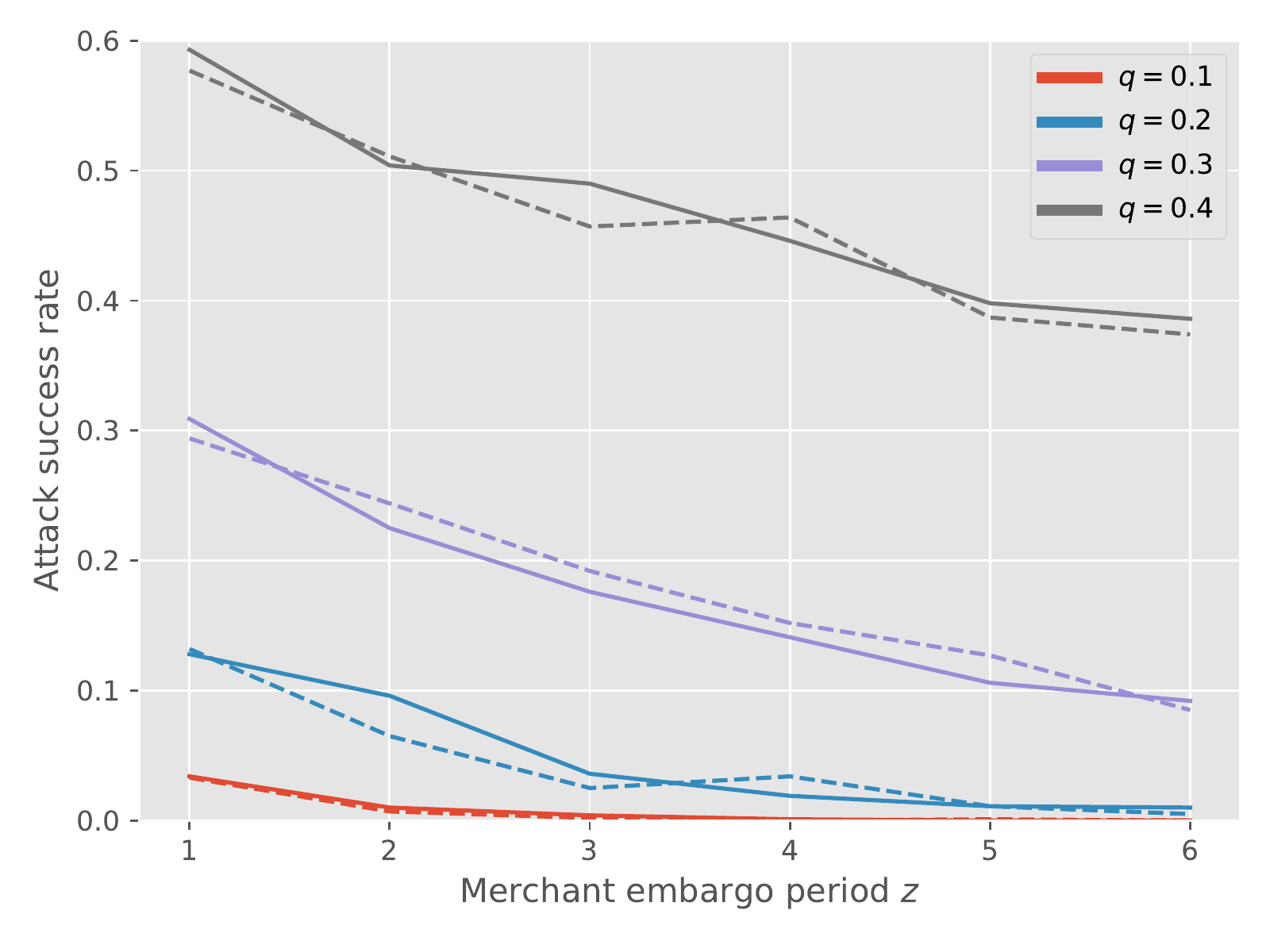}
  \caption{Probability (dependent axis) that an attacker, having fraction $q$ of the total hash rate (individual lines), succeeds in a doublespend attack when merchants impose an embargo period (independent axis) of $z$ blocks. The solid and dashed lines indicate success probability in the Bitcoin and Radium protocols, respectively. Each point represents the success frequency over 1000 trials.}
  \label{fig:ds_prob}
\end{figure}

Let $Z_j$ be a random variable representing the block time using Bobtail with parameter $j$; i.e., there are $j$ proofs per block. It has been shown~\cite{Bissias:2020} that the improvement in variance relative to Bitcoin is given by
\begin{equation}
\frac{Var[Z_j]}{Var[X]} = \frac{8j+4}{6(j^2+j)}.
\end{equation}
Finally, we can determine the value of $j$ for which Bobtail's improvement in variance is equivalent to RTT with $k=2$ by solving
\begin{equation}
\frac{4}{\pi} - 1 = \frac{8j+4}{6(j^2+j)}.
\end{equation}
Solving for $j$ we have
\begin{equation}
j = \frac{7 \pi - 12 + \sqrt{144 -72 \pi + 25 \pi^2}}{6(4 - \pi)} > 4.
\end{equation}

Yet despite the fact that the reduction in variance for Radium is roughly equivalent to Bobtail with $j = 4$, it turns out that Radium has the same susceptibility to doublespend attacks as does Bitcoin. Figure~\ref{fig:ds_prob} shows the result of a mining simulator that we ran for both Bitcoin (solid lines) and Radium (dashed lines). Each curve represents a different attacker hash power, ranging from 10\% up to 40\% of the total. Points along each curve correspond to the embargo period imposed by the coin receiver, a merchant for example. For an embargo period of length $z$, the merchant will not release goods purchased with a transaction in block $i$ until $z$ additional blocks have been mined after it. The figure shows that there are negligible differences between attacker success probability when comparing Bitcoin to Radium. 

The results of our simulation suggest that there might be something fundamental about the doublespend protection afforded by protocols that use just one PoW sample per block. We formalize this conjecture below, but leave investigation to future work.

\begin{myconj}
Mining a block under PoW amounts to sampling a sufficiently low statistic from a known distribution. For example, in Bitcoin, the statistic is a single exponential random variable. Let $K$ be any mining statistic on a single sample per block, i.e. a single random variable is sampled once. And assume that $K$ is fair in the sense that a miner with fraction $x$ of the hash rate receives fraction $x$ of the rewards in expectation. Then the doublespend protection offered by a protocol using $K$ is no better than that offered by a protocol using an exponential random variable for its statistic.
\end{myconj}

\section{Conclusion}

We have identified and analyzed a critical vulnerability in the real-time block rate targeting protocol (RTT). To mitigate this vulnerability, we introduced Radium, a refinement of RTT. Like RTT, Radium  offers less variable block times, a more responsive DAA, and thwarts the chain death spiral that threatens minority hash rate blockchains. We have also shown that Radium maintains Bitcoin's robustness to the doublespend attack as well as its low orphan rate.

\urlstyle{sf}
\pagestyle{plain}
{\footnotesize \bibliographystyle{acm}
\bibliography{references}}

\end{document}